# Biophysical and biomechanical properties of cartilage

**Enrico Catalano[1]**

[1] University of Oslo (UiO), Oslo, Norway.

**Abstract:** Cartilage is a connective tissue that covers the surfaces of bones in joints and provides a smooth gliding surface for movement. It is characterized by specific biophysical properties that allow it to withstand compressive loads, distribute mechanical forces, and maintain tissue integrity. The biophysical properties of cartilage are primarily determined by its extracellular matrix, which is composed of collagen fibers, proteoglycans, and water. The collagen fibers provide tensile strength, the proteoglycans provide compressive resistance, and the water content provides lubrication and shock absorption. The potential for greater knowledge of cartilage function through refinement and engineering-level understanding could inform the design of interventions for cartilage dysfunction and pathology. The aim is to assist to present basic principles of cartilage modeling and discussing the underlying physics and assumptions with relatively simple settings, and also it presents the derivation of multiphase cartilage models that are consistent with the discussions. Furthermore, modern developments align the structure captured in the models with observed complexities. The interactions between these components and the surrounding tissues regulate cartilage biomechanics and contribute to its ability to resist damage and repair itself. Alterations in the biophysical properties of cartilage can lead to degenerative joint diseases such as osteoarthritis, highlighting the importance of understanding cartilage structure and function for the development of effective therapeutic strategies.

## 1. Introduction

Cartilage is a specialized connective tissue that is essential for the proper functioning of joints in the body. It is characterized by its flexible and elastic properties that provide support and cushioning to the bones, allowing for smooth movement and shock absorption. Cartilage is composed of chondrocytes (cells), collagen and elastin fibers, and a matrix of water, proteoglycans, and glycosaminoglycans (GAGs). Its unique composition and structure give it its distinct properties, including high tensile strength, compression resistance, and durability. Understanding the properties of cartilage is crucial for developing effective treatments for common joint disorders, such as osteoarthritis and rheumatoid arthritis.. Articular cartilage is found at opposing bone surfaces in joints and is a remarkable tissue, characterised by extremes of physiological structure and mechanical function: it lacks vasculature, lymphatics and nerves yet exhibits tribological properties that surpass engineering standards. With extremes of performance and loading over a whole lifetime, it is no surprise to find that articular cartilage function is both mechanically complex, and prone to degeneration and pathology.

The mechanical performance of cartilage is underpinned by its structure, which has been extensively documented. The major constituents of cartilage include an anisotropic and heterogeneous matrix of predominantly type II collagen, intermeshed with high molecular weight proteoglycans, mainly aggrecans, immersed in an interstitial fluid containing numerous physiological electrolytes. Near the interface with bone, the collagen matrix is oriented predominantly perpendicular to the bone when averaged at the mesoscale, with average fibril orientation rotating on moving up through the cartilage until the fibrils are mainly parallel to the articulating surface. From theoretical studies, this mesoscale architecture is understood to ameliorate tissue stresses under load, and enables a smooth transition of functionalities derived from microstructural effects.

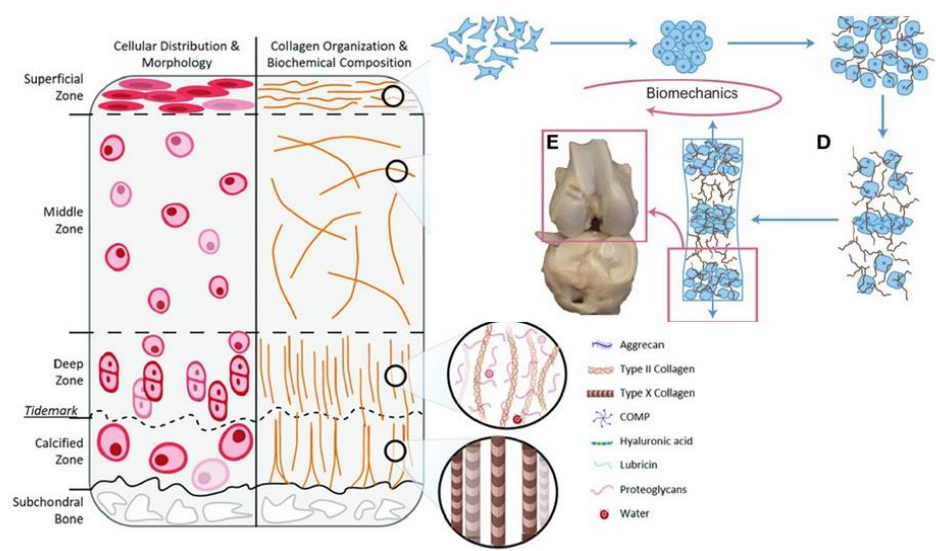

**Figure 1.** Biostructure of cartilage

**Biomechanical properties of cartilage and chondrocytes**

The main mechanical properties of cartilage are related to composition: cartilage is a connective tissue that is composed of chondrocytes, extracellular matrix, and collagen fibers, elasticity: cartilage has a high degree of elasticity, which allows it to deform under pressure and return back to its original shape, density: cartilage has a low density compared to bone, making it a soft and pliable tissue, lubrication: cartilage is coated with synovial fluid, which helps to lubricate the joint and reduce friction between bones, avascularity: Cartilage is avascular, meaning it does not have its own blood supply. Nutrients and oxygen are delivered by diffusion from surrounding tissues.

The chondrocytes have the following properties: shape: Chondrocytes are round or oval-shaped cells that are found in cartilage tissue, matrix synthesis: chondrocytes produce and secrete the extracellular matrix of cartilage, which consists of proteoglycans and collagen fibers, division: chondrocytes have a low rate of division and can only proliferate under certain conditions, diffusion rate: chondrocytes have a low diffusion rate due to the dense extracellular matrix surrounding them, nutrient supply: chondrocytes rely on diffusion of nutrients and oxygen from surrounding tissues as they are also avascular like cartilage. From this point of view the biophysical characterization of properties of cartilage.

**Determination of osmotic parameters of cartilage**

The relation between the equilibrium volume of an articular cartilage chondrocyte and the osmolarity of the solution is described by the Boyle-Van't Hoff equation,

$$V/V_{\text{iso}} = (1 - V_{\text{b}}/V_{\text{iso}})M_{\text{iso}}/M + V_{\text{b}}/V_{\text{iso}}, \tag{1}$$

where V is the equilibrium volume at osmolarity M, $V_{iso}$ is the volume under isotonic conditions ($M_{iso}$), and $V_b$ is the osmotically inactive volume of the chondrocyte. In determining the osmotically inactive volume of the chondrocyte, the volumes of the measured articular cartilage chondrocytes can be normalized to their respective isotonic volumes ($V/V_{iso}$) as the y axis, with the reciprocal of normalized osmolarity ($M/M_{iso}$) as the x axis. The y intercept of the linear regression line shows the osmotically inactive fraction ($V_b/V_{iso}$) of the chondrocyte, which was then used to determine the osmotically inactive volume of the chondrocyte.

To determine the hydraulic conductivity of the cell membrane, the rate of water transport across the cell membrane can be described as

$$dV/dt = L_P A R T (M_i - M_e)$$
(2)

where $L_P$ is the membrane's hydraulic conductivity, A is the surface area of the plasma membrane, R is the universal gas constant, T is the absolute temperature, $M_e$ is the extracellular osmolarity, and $M_i$ is the intracellular osmolarity, which can be calculated from Eq. (1). The surface area of the plasma membrane was assumed to remain constant during the volume change; this implies that the membrane did not shrink or stretch, but rather folded during cell shrinkage and unfolded during swelling. $L_P$ was determined by fitting the experimental data to Eqs. (1) and (2), using the Levenberg-Marquardt optimization scheme [10], [28], [34], [36]. The Levenberg–Marquardt optimization scheme was used to seek the optimal LP that maximized $R^2$, the coefficient of determination defined as

$$R^2 = 1 - \frac{\sum(y_i - y_{fit,i})^2}{\sum(y_i - \bar{y})^2}$$
(3)

where $y_i$ are the experimental data, $y_{fit,i}$ are the corresponding model predictions, and $\bar{y}$ is the arithmetic mean of the experimental data.

The Arrhenius temperature dependence of the membrane hydraulic conductivity is,

$$L_P = L_{Pg} \exp\left[\frac{E}{R}\left(\frac{1}{T_{ref}} - \frac{1}{T}\right)\right]$$
(4)

where LP is the membrane hydraulic conductivity at absolute temperature T, LPg is membrane hydraulic conductivity at reference temperature, R is the universal gas constant, Tref is the reference temperature (273.15 K), and E is the activation energy of water transport.

**Biophysical description of cartilage**

Cartilage is a specialized connective tissue that plays a crucial role in providing structural support and shock absorption in the body. It is found in several locations, including the joints, ears, and nose. The biophysical properties of cartilage are critical in understanding its function and behavior, especially in the context of diseases such as osteoarthritis. The key biophysical properties of cartilage and the equations that describe them are as follows:

<u>Elasticity</u>

Cartilage is a highly elastic tissue that can deform under mechanical stress and then return to its original shape. The elastic behavior of cartilage can be described using the Hooke's law equation:

$$\sigma = E\varepsilon$$

where σ is the stress applied to the cartilage, E is the modulus of elasticity (also known as Young's modulus), and ε is the strain or deformation of the tissue. The modulus of elasticity is a measure of the stiffness of the tissue and is dependent on the composition and structure of the cartilage.

<u>Viscoelasticity</u>

In addition to its elastic properties, cartilage also exhibits viscoelastic behavior, meaning that it can deform under stress and then exhibit a time-dependent recovery. The viscoelastic behavior of cartilage can be described using the Kelvin-Voigt model, which combines the elastic behavior of Hooke's law with the viscous behavior of Newton's law:

$$\sigma = E\varepsilon + \eta(d\varepsilon/dt)$$

where η is the viscosity of the tissue and dε/dt is the rate of deformation. The Kelvin-Voigt model is useful in describing the complex behavior of cartilage under cyclic loading, such as during joint movement.

<u>Permeability</u>

Cartilage is a porous tissue that allows for the transport of nutrients and waste products between the tissue and the synovial fluid that surrounds it. The permeability of cartilage can be described using Darcy's law:

$$Q = -KA(dP/dx)$$

where Q is the fluid flow rate, K is the hydraulic conductivity of the tissue, A is the cross-sectional area, dP/dx is the pressure gradient across the tissue, and - sign indicates that fluid flows from high to low pressure. The hydraulic conductivity is a measure of how easily fluid can flow through the tissue, and is dependent on the size and shape of the pores in the tissue.

<u>Friction</u>

In joints, cartilage functions to reduce friction between the bones during movement. The frictional properties of cartilage can be described using Coulomb's law of friction:

$$Ff = \mu Fn$$

where Ff is the frictional force, μ is the coefficient of friction, and Fn is the normal force between the two surfaces in contact. The coefficient of friction is a measure of how effectively the cartilage reduces friction between the bones, and is dependent on the composition and structure of the tissue.

The biophysical properties of cartilage are crucial in understanding its function and behavior. The elasticity, viscoelasticity, permeability, and frictional properties of cartilage can be described using several equations, such as Hooke's law, the Kelvin-Voigt model, Darcy's law, and Coulomb's law of friction. These equations provide a framework for studying the behavior of cartilage under different loading and environmental conditions, and can be used to develop new strategies for treating cartilage-related diseases.

**Cartilage Boundary conditions**

The boundary conditions for the cartilage, modelled as a bi- or tri- phasic material, are more vexatious to specify. Firstly, note that the solid phase moves with the cartilage boundary so characteristics of the solid phase velocity do not enter the cartilage domain and hence no boundary conditions are required for the equation governing the solid phase mass balance. However, if there is a flux of fluid into the cartilage, then a flux boundary condition for the fluid phase mass balance hyperbolic partial differential equation is required. In addition, there are numerous further interfacial conditions for the finger cartilage in this model.

Cartilage-Bone: At the lower, assumed static, bone zero velocity for both fluid and solid phase is imposed,

$$\boldsymbol{v}^{\mathrm{f}} = \boldsymbol{v}^{\mathrm{s}} = \boldsymbol{0}$$

On the upper bone-cartilage interface continuity of both velocity and stress for both phases are imposed, noting that the upper bone can move in response to the applied force, as represented by the pressure p*. Thus firstly,

$$\boldsymbol{v}^{\text{f}} = \boldsymbol{v}^{\text{s}} = \boldsymbol{v}^{bone}$$

In addition, on the upper bone-cartilage interface the stress from the synovial fluid must be distributed between the solid and fluid cartilage phases. In the absence of detailed empirical guidance, this is implemented via the area fractions of solid and fluid phase presenting at the bone-cartilage interface, though approximating collagen fibres as cylinders, these area fractions can readily be shown to be equivalent to volume fractions. For instance near bone, collagen fibres are essentially perpendicular to the interface, and for an array of N vertical cylinders of radius a, the ratio of surface area to volume fractions with a region of area A and height H is given by

$$\left(\frac{N\pi a^2}{A}\right)\left(\frac{HA}{NH\pi a^2}\right) = 1$$

Hence we have

$$\boldsymbol{\sigma}^s \cdot \boldsymbol{n} = -p^* \boldsymbol{n} \phi^s, \quad \boldsymbol{\sigma}^f \cdot \boldsymbol{n} = -p^* \boldsymbol{n} \phi^f$$

Cartilage-synovial fluid. At the cartilage-synovial interface, continuity of stress is imposed for both cartilage phases, again with area fractions taken as equivalent to volume fractions. Hence

$$\boldsymbol{\sigma}^s \cdot \boldsymbol{n} = \boldsymbol{\sigma}^{Synovial} \cdot \boldsymbol{n} \phi^s, \quad \boldsymbol{\sigma}^f \cdot \boldsymbol{n} = \boldsymbol{\sigma}^{Synovial} \cdot \boldsymbol{n} \phi^f$$

Additional conditions are required. Continuity of normal synovial fluid and normal solid cartilage velocity is imposed, $\boldsymbol{v}^s \cdot \boldsymbol{n} = \boldsymbol{v}^{Synovial} \cdot \boldsymbol{n}$, representing a kinematic condition delimiting the evolution of the interface. We also have the cartilage fluid normal velocity and normal synovial fluid velocities are related by mass balance across the interface; assuming the cartilage fluid and synovial fluid have the same density, this is simply the constraint $\boldsymbol{v}^{Synovial} \cdot \boldsymbol{n} = \phi^f \boldsymbol{v}^f \cdot \boldsymbol{n}$.

At this stage only tangential fluid velocities and tangential solid displacements, or alternatively tangential stresses, have not been subjected to boundary constraints.

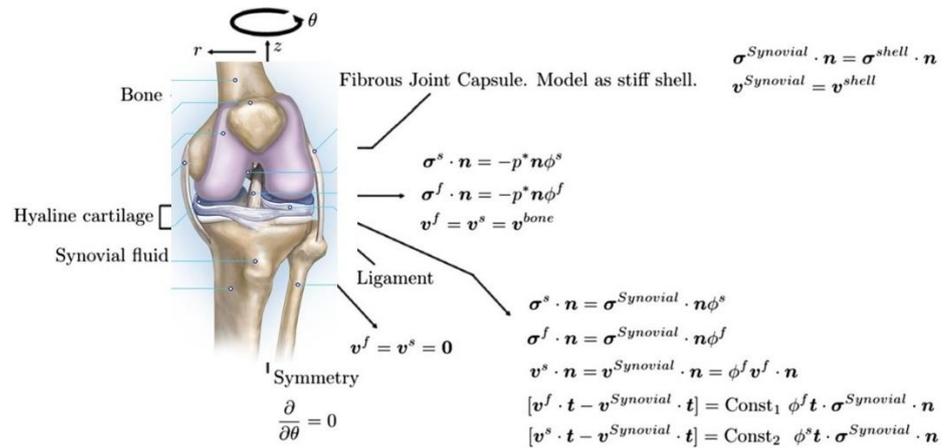

**Figure 2.** Biomechanical properties of cartilage

### Multiphase models of cartilage

This model details and derives the formulation of a core biphasic model, representing the solid and fluid phases of cartilage, before introducing the effect of osmotic pressures and examining boundary conditions to represent cartilage structure in more extensive detail, such as the inclusion of collagen fibre anisotropies.

### Biphasic poroelastic models

The biphasic model, developed by Mow and colleagues, is the most basic poroelastic cartilage framework. It consists of two components – the solid matrix and interstitial fluid – which fill space. The model is reduced to the classical Biot model for infinitesimal strain, as described by Bowen in 1980. In this model, continuum fluid dynamics is used for the interstitial phase fluid dynamics, which differs from traditional frameworks such as those proposed by Mow et al. in 1980, Lai et al. in 1991, and Huyghe and Janssen in 1997. However, the resulting framework is similar to the standard biphasic model, and any differences between the two can be tested.

### Triphasic and quadphasic poroelastic models of cartilage

The triphasic modelling framework can describe cartilage structure by incorporating an ionic phase, which expanded biphasic models. As a result, complex models with simpler constitutive relationships were created, enabling a closer connection between observed processes and modelling representation. While the solid phase of tissue is typically represented by a homogeneous, isotropic, linearly elastic material experiencing infinitesimal strain, triphasic treatments can typically be related to biphasic models after the properties of aggrecans undergo minimal changes (Ateshian et al., 2004).

A quadphasic description of cartilage tissue can be modelled, which separates cations and anions to more accurately capture swelling (Huyghe and Janssen, 1997) to provide generalisations for an arbitrary number of ion species (Gu et al. (1998) and Ateshian (2007)). A most comprehensive and generalised model for mixture theory and rational thermodynamics in soft tissue modelling such as cartilage is available, featuring a description of growth or reactions among constituents for cartilage and soft tissue modelling.

### Mass and momentum balance of cartilage

The properties of cartilage are regulated by microscale incompressibility with the mass densities $\varrho^f$ and $\varrho^s$ for the fluid and solid phases, that are related to their volume fractions $\phi^f$, $\phi^s$ by the constant, true, density of the phases, via $\rho_T^\beta \phi^\beta = \rho^\beta$, with $\beta \in \{f, s\}$, and where the T subscript denotes true density. In addition, demanding the material fills all space yields

$$1 = \phi^f + \phi^s. \tag{1}$$

furthermore, velocity fields of the fluid phase are present, denoted by $v^f$, and the displacement of solid phase material points $u^s(X, t) = x(X, t) - X$, where $x(X, t)$ maps material points, $X$ in the reference configuration to points $x$ in the inertial frame at a given time, t; these represent 6 unknown scalar fields. The constraint (1) can be enforced using a Lagrange multiplier, denoted p, via the Coleman-Noll procedure (Coleman and Noll, 1963). Nine macroscale scalar unknowns, $\{u^s, v^f, p, \phi^f, \phi^s\}$ can be modulated, and hence 9 scalar equations are required, of which constraint (1) is one. Two more scalar equations arise from mass balances, with a further six from the balance of momentum.

With the assumptions of mixture theory for porous media (Bowen, 1980), equivalent to a multiphase model given constituent incompressibility, we have the macroscale mass balances

$$\frac{\partial \rho^\beta}{\partial t} + \frac{\partial}{\partial x_i}\left(\rho^\beta v_i^\beta\right) = 0, \quad \beta \in \{f, s\},$$

where $v^s := Du^s / Dt$, $\rho_T^\beta \phi^\beta = \rho^\beta$ and where $\partial/\partial t$ holds $x$, a point in the inertial reference frame, fixed in contrast to the derivative D / D$_t$, which is relative to a material point, denoted $X$, fixed in the reference configuration of the solid phase.

In formulating the momentum balance equations, we assume the system is isothermal and with negligible inertia, as motivated in the ESM, Section 2. Further noting, as implicit in the mixture theory framework, that all stress gradients and drag forces are defined per unit volume of tissue, mixture theory postulates the following form of the macroscale equations

$$\frac{\partial \sigma_{ij}^f}{\partial x_j} + q_i^f = 0, \quad \frac{\partial \sigma_{ij}^s}{\partial x_j} + q_i^s = 0, \quad q^f = -q^s = \gamma\left(v^s - v^f\right) + \Delta q.$$

Here $\sigma^f, \sigma^s$ are the Cauchy stress tensors for the elastic and solid phases and $q^f$ is the stress exerted on the fluid phase by the solid phase, with $q^f = -q^s$ by Newton's Third Law. A standard choice for this term is $\gamma(v^s - v^f)$, with $\gamma > 0$ constant. However, this immediately discounts how the detailed structure impacts the relative forces between the two phases and, once more, information about cartilage structure is lost. Finally, an additional drag term for biophysical characterization, $\Delta q$ to enforce thermodynamic consistency is fundamental.

**Elasticity and Viscoelasticity of cartilage**

One of the key properties of cartilage is its elasticity, which allows it to deform under load and return to its original shape when the load is removed. This property is due to the presence of water and the high concentration of proteoglycans in the cartilage matrix, which provide resistance to compression. Cartilage is also viscoelastic, meaning that it exhibits both viscous and elastic behavior under load. This property is due to the highly hydrated nature of the tissue, which makes it able to deform and absorb shock under a sudden or sustained load. The viscoelastic properties of cartilage are important in joints, where they allow the tissue to absorb shock and distribute load across a wider area. This helps to protect the underlying bone and prevent wear and tear.

**Mechanical Strength**

Another important property of cartilage is its mechanical strength, which is crucial for maintaining its function as a load-bearing tissue. Cartilage is able to resist compressive and tensile forces due to the collagen fibers embedded in the matrix.

The mechanical strength of cartilage varies depending on its location in the body. For example, the articular cartilage found in joints is highly compressed and must withstand high shearing and compressive stresses. In contrast, the ear cartilage is less mechanically strong due to its lower collagen content and higher water concentration.

**Friction and Lubrication**

Cartilage in joints also has important biophysical properties related to friction and lubrication. Under normal conditions, the cartilage surface is smooth and covered with a layer of synovial fluid that provides lubrication and reduces friction between the joint surfaces. However, when the cartilage is damaged or degraded, it can become rough and unstable, leading to increased friction and wear. This can cause pain, inflammation, and joint deterioration, leading to conditions such as osteoarthritis.

**Conclusions**

Overall, the biophysical properties of cartilage are crucial for its function and performance as a load-bearing tissue. Understanding these properties is important for the development of new treatments for joint and cartilage disorders, as well as for the design of prosthetic materials that mimic the biomechanical properties of natural cartilage. biophysics plays a critical role in understanding the structure and function of cartilage. Cartilage is a complex tissue that has unique mechanical and physical properties. These properties are directly related to the composition and organization of its extracellular matrix, which consists mainly of collagen, proteoglycans, and water. Biophysical studies have revealed how these components interact to create the mechanical properties of cartilage and how external factors such as loading affect its behavior. Additionally, biophysical techniques like molecular dynamics simulations, X-ray diffraction, and atomic force microscopy have provided new insights into the molecular-level mechanisms that govern cartilage behavior, leading to the development of new therapeutic approaches for treating cartilage damage and degeneration. Thus, biophysics is an essential tool for understanding the biophysical principles underlying cartilage function and for developing strategies for treating cartilage-related diseases.

**References**

- K. Athanasiou, E. Darling, J. Hu, G. DuRaine, A. Hari Reddi. Articular Cartilage. CRC Press, Boca Raton, FL (2013)
- F.P. Rojas, M.A. Batista, C.A. Lindburg, D. Dean, A.J. Grodzinsky, C. Ortiz, L. Han. Molecular adhesion between cartilage extracellular matrix macromolecules. Biomacromolecules, 15 (3) (2014), pp. 772-780
- N. Broom, A. Oloyede. Experimental-Determination Of The Subchondral Stress-Reducing Role Of Articular-Cartilage Under Static And Dynamic Compression. Clin. Biomech., 8 (2) (1993), pp. 102-108
- V. Mow, S. Kuei, W. Lai, C. Armstrong Biphasic creep and stress relaxation of articular cartilage in compression: theory and experiments. J. Biomech. Eng., 102 (1) (1980), pp. 73-84
- Mow, V., Kwan, M., Lai, W., Holmes M., 1986. A finite deformation theory for nonlinearly permeable soft hydrated biological tissues. In: G. Schmid-Schonbein, S.-Y. Woo, B. Zweifach (Eds.), Frontiers in Biomechanics, Springer, New York, pp. 153–179.

- Wilson et al., 2007 W. Wilson, J. Huyghe, C. Van Donkelaar Depth-dependent compressive equilibrium properties of articular cartilage explained by its composition. Biomech. Model. Mechanobiol., 6 (1–2) (2007), pp. 43-53